\documentclass[reprint,amsmath,amssymb,aps,twocolumn,superscriptaddress]{revtex4-2}
\usepackage{graphicx}
\usepackage{dcolumn}
\usepackage{bm}
\usepackage{float}
\usepackage{amssymb}
\usepackage{color}
\usepackage{multirow}
\usepackage{stmaryrd}
\usepackage[colorlinks,linkcolor=blue,urlcolor=blue,citecolor=blue]{hyperref}

\begin{document}


\title{Unveiling High-Energy Ultrafast Responses Driven by Fermi-Level Gap Opening in Charge-Density Waves}

\author{Qiong Wu}
\thanks{Co-first author}
\email{qwu@cqu.edu.cn}
\affiliation{School of Physics and Center of Quantum Materials and Devices, Chongqing University, Chongqing 401331, China}
\affiliation{International Center for Quantum Materials, School of Physics, Peking University, Beijing 100871, China}

\author{Tianchen Hu}
\thanks{Co-first author}
\email{tchu@pku.edu.cn}
\affiliation{International Center for Quantum Materials, School of Physics, Peking University, Beijing 100871, China}

\author{Junhan Huang}
\affiliation{International Center for Quantum Materials, School of Physics, Peking University, Beijing 100871, China}

\author{Yongjie Liu}
\affiliation{College of Materials Science and Engineering, Chongqing University of Arts and Sciences, Chongqing 402160, China}

\author{Fei Sun}
\affiliation{Max Planck Institute for Chemical Physics of Solids, Dresden 01187, Germany}
\affiliation{Quantum Science Center of Guangdong-Hong Kong-Macao Greater Bay Area, Shenzhen, 518045, China}

\author{Shuxiang Xu}
\affiliation{Center for High Pressure Science and Technology Advanced Research, Beijing 100193, China}

\author{Li Yue}
\affiliation{Beijing Academy of Quantum Information Sciences, Beijing 100193, China}

\author{Zijian Lin}
\affiliation{International Center for Quantum Materials, School of Physics, Peking University, Beijing 100871, China}

\author{Xinyu Zhou}
\affiliation{International Center for Quantum Materials, School of Physics, Peking University, Beijing 100871, China}

\author{Tao Dong}
\affiliation{ Tsung-Dao Lee Institute, School of Physics and Astronomy, Shanghai Jiao Tong University, Shanghai 201210, China}
\affiliation{International Center for Quantum Materials, School of Physics, Peking University, Beijing 100871, China}

\author{Nan-Lin Wang}
\email{nlwang@sjtu.edu.cn}
\affiliation{ Tsung-Dao Lee Institute, School of Physics and Astronomy, Shanghai Jiao Tong University, Shanghai 201210, China}
\affiliation{International Center for Quantum Materials, School of Physics, Peking University, Beijing 100871, China}
\affiliation{Beijing Academy of Quantum Information Sciences, Beijing 100193, China}
\affiliation{Zhangjiang Institute for Advanced Study, Shanghai Jiao Tong University, Shanghai, 201210, China}

\date{\today}

\begin{abstract}

Ultrafast pump–probe spectroscopy is indispensable for investigating interactions, collective excitations, and nonequilibrium dynamics in superconductors and correlated electron systems. However, how a small gap opening near Fermi surface affects the spectral changes in the high-energy interband transition region remains elusive. To address this issue, we perform systematic studies of carrier relaxation dynamics in the prototype charge-density-wave (CDW) material EuAl$_4$. We find that the CDW gap shrinks substantially under optical pumping. This band renormalization strongly modulates interband transitions associated with the gap edge, inducing a pronounced transient reflectivity change at the corresponding probe energies. These results establish a direct link between low-energy order-parameter dynamics and high-energy spectral responses, providing a general framework for interpreting transient signals in gapped quantum materials.
 
\end{abstract}

\pacs{Valid PACS appear here}
\maketitle


Pump-probe spectroscopy is a cornerstone technique in ultrafast science. By exciting elementary and collective modes in condensed matter with an intense femtosecond pump pulse and subsequently interrogating the system with a time-delayed, weaker probe pulse, this method captures the temporal evolution of nonequilibrium states with exceptional precision\cite{Giannetti2016,Kirilyuk2010,Chen2017,Wang2021}. It has proven particularly powerful in studying symmetry broken systems with formation of single-particle energy gaps, such as superconductors and charge density wave (CDW) materials\cite{Demsar1999,Demsar1999a,Perfetti2007,Yusupov2008,Chia2010,Torchinsky2010}. In these ordered phases, there is a peculiar carrier relaxation channel, whose amplitude diminishes and lifetime diverges as the critical point is approached—a hallmark behavior well described by the Rothwarf-Taylor (RT) model, which captures the recombination dynamics of photoexcited quasiparticles across an energy gap\cite{Rothwarf1967,Demsar1999}.

A fundamental question, however, remains unresolved: how can a low-energy gap on the meV scale exert such profound modulation on the transient reflectivity probed at eV energies? Conventionally, optical responses in the visible-to-near-infrared range are attributed to interband transitions far away from the Fermi level and are expected to be largely insensitive to low-energy order parameters\cite{Dressel2002}, consistent with the temperature-independent steady-state reflectivity observed in this spectral window\cite{Hu2008,Li2008,Zhou2021}. This expectation stands in stark contrast to the robust gap-induced transient signals routinely measured in pump-probe experiments. To resolve this paradox, it is essential to investigate the transient reflectivity over a broad energy range, thereby capturing the full impact of gap opening on the ultrafast dynamics. Single-color or two-color approaches lack the necessary spectral coverage; broadband probing with a supercontinuum white-light source is therefore essential to capture the full energy-dependent fingerprint of the ultrafast response\cite{Jnawali2020,Obergfell2020}.

\begin{figure*}[t]
	\centering
	\includegraphics[width=14cm]{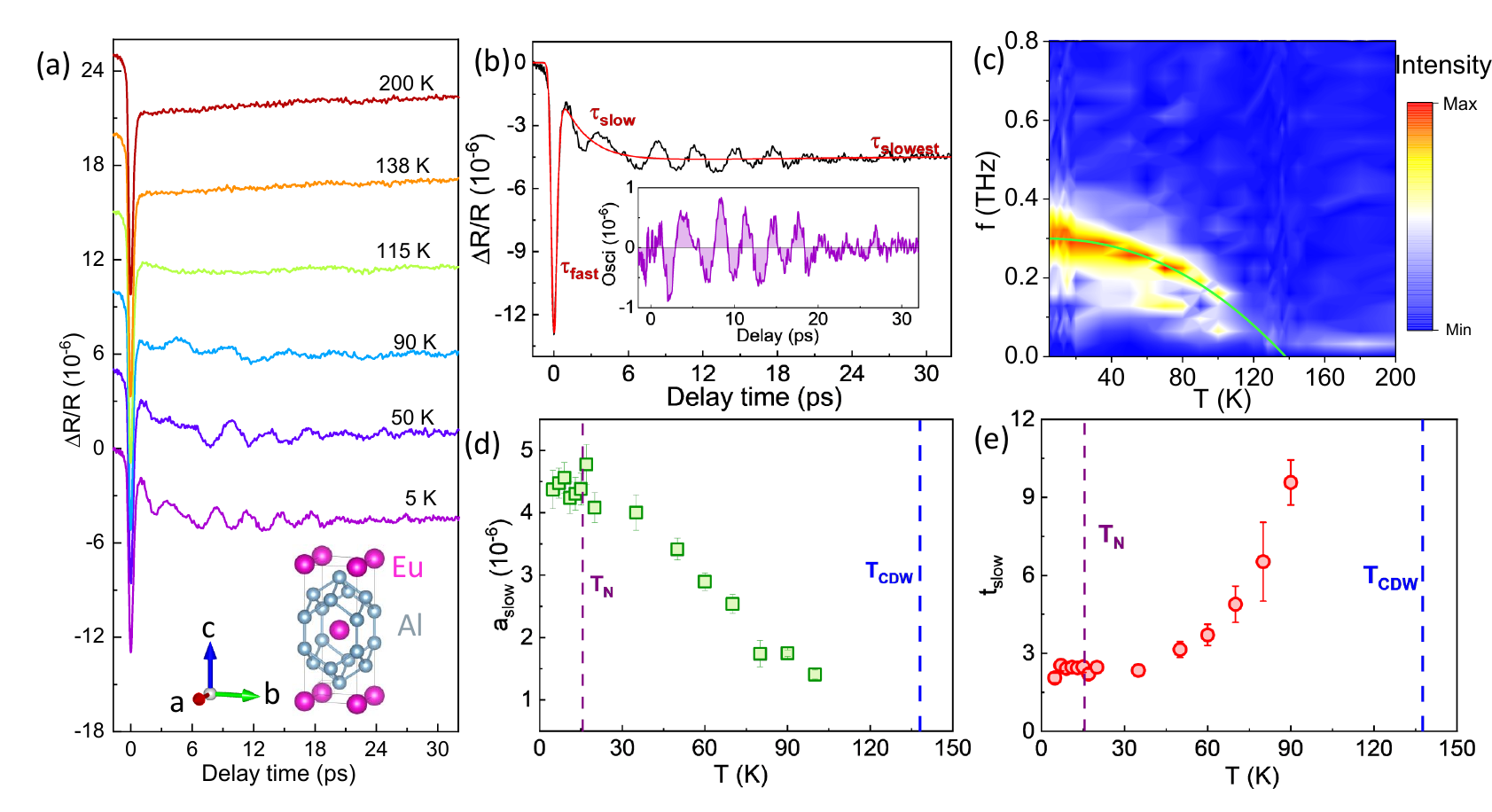}\\
	\caption{\textbf{The temperature-dependent ultrafast relaxation dynamics of EuAl$_4$.}  (a) Ultrafast dynamics at several typical temperature points, showing the temperature dependence of oscillations and a positive component. The data are vertically offset for clarity. The inset illustrates the crystal structure of EuAl$_4$. (b) Fitting of the 5K data using the convolution of a three-exponential function with uniform Gaussian function. The inset shows the pure oscillatory component remaining after subtracting the fitted background. (c) Two-dimensional color map of the Fast Fourier transition of pure oscillation, showing tendency of frequency softening; the green curve is guide to the eyes. Temperature dependence of (d) amplitude and (e) lifetime of the positive slow component are fitted according to panel (b).}\label{Fig:1}
\end{figure*}

Here we investigate the broadband quasiparticle dynamics across the CDW transition--a prototypical example of spontaneous symmetry breaking\cite{Gruener1988,Rice1975}--in EuAl$_4$(Section I in \cite{Wu2026}), a CDW system with a transition temperature $T_{CDW}$ $\approx$ 140 K\cite{Zhang2013,Araki2013,Nakamura2015,Stavinoha2018,Shang2021,Takagi2022,Meier2022,Gen2023,Vibhakar2024,Wang2024,Yang2024,Ni2024,Eaton2024,Sukhanov2025}, with an incommensurate modulation vector along the c axis, and exotic magnetic order emerging at lower temperatures. This material was selected because its key optical response fall squarely within the spectral window of our supercontinuum probe, providing an ideal platform for energy-resolved ultrafast spectroscopy. Dual-color pump-probe measurements first confirm the applicability of the RT framework: a relaxation component emerges or becomes enhanced in the CDW phase with a lifetime that diverges near $T_{CDW}$, accompanied by a pronounced amplitude mode (AM) response. Broadband probing then resolves the distinct spectral fingerprints of each dynamical component. Among them, the CDW-associated relaxation component exhibits an energy-dependent dip-hump lineshape around 2 eV, this arises from the modulation of gap-edge-associated transitions by the shrinking CDW gap. This identification not only clarifies the microscopic mechanism underlying visible-to-near-infrared pump-probe responses, but also establishes a generalizable analytical paradigm applicable to other gapped quantum phases. Furthermore, by modeling the energy-resolved dynamics, we extract both the transient suppression of the CDW gap and the relaxation trajectory of the lattice distortion normal coordinate following photoexcitation. These results underscore the unique capability of broadband ultrafast spectroscopy to disentangle coupled electronic and structural degrees of freedom in nonequilibrium quantum materials.

\begin{figure*}[t]
	\centering
	\includegraphics[width=14 cm]{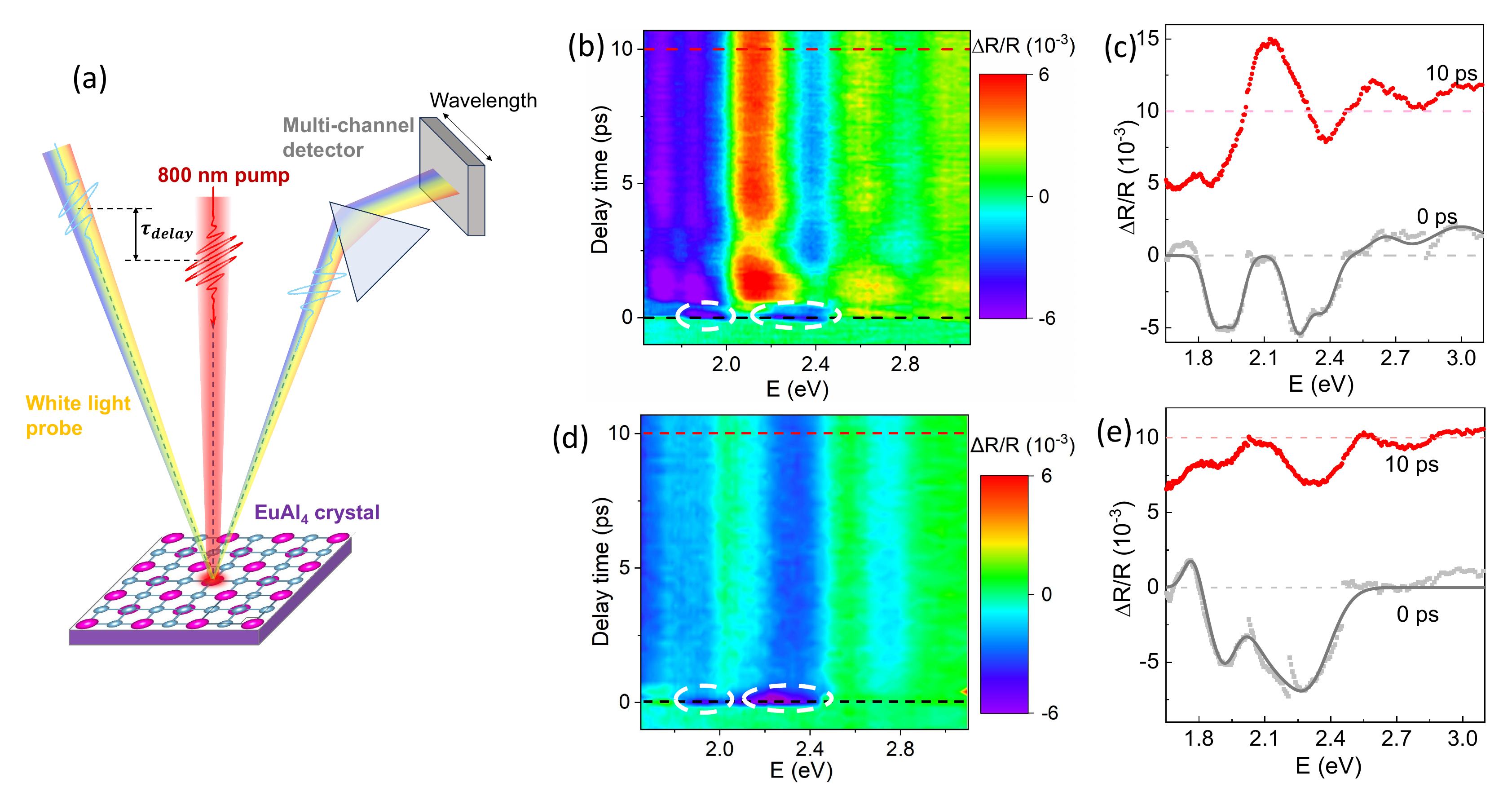}\\
	\caption{\textbf{Broadband transient differential reflectivity spectra of EuAl$_4$ in the CDW  and normal state.}
	(a) Configuration of broadband reflectance measurement. The probe beam is broadband white light, which can be separated into different wavelengths through diffraction or refraction after reflection. By measuring the pump and no-pump light intensities at various positions of a multi-channel detector, one can obtain the transient reflectivity as a function of probe wavelength (or equivalently, energy) and delay time. (b) Two-dimensional color map of the $\Delta R/R$ spectrum at 20 K. The two dashed white circles highlight the short-lived features at 0 ps. (c) $\Delta R/R$ spectra at delay times of 0 ps and 10 ps at 20 K. (d) Two-dimensional color map of $\Delta R/R$ spectrum at 150 K. (e) $\Delta R/R$ spectra at delay times of 0 ps and 10 ps at 150 K. The data in panel (c) and (e) are vertically offset for clarity.}\label{Fig:2}
\end{figure*}

Having verified the sample quality (Section II and III in \cite{Wu2026}), we first perform two-color pump-probe measurements to determine whether the opening of the CDW gap in EuAl$_4$ introduces an additional quasiparticle relaxation channel (see section IV, V, VI in \cite{Wu2026}). At 5 K, the ultrafast relaxation dynamics exhibits three components: (i) a dominant negative fast component with a lifetime of $\sim$ 0.2 ps; (ii) a positive slow component with $\sim$ 2 ps lifetime; and (iii) a negative slowest component persisting beyond 30 ps (Fig. 1(a,b)). Low-frequency oscillations are superimposed on this relaxation background. Both the slow component and oscillations weaken with increasing temperature, disappearing above 138 K, indicating their close association with the CDW phase (Fig. 1(a)). We employ a convolved triple-exponential function to extract component amplitudes and lifetimes(Section IV in \cite{Wu2026}):
\begin{equation}
\begin{aligned}
\Delta R/R (t) = G(t) \ast \left( \sum_{i=1}^{3} a_i \cdot \exp(-t/\tau_i) \right), \
\end{aligned}
\end{equation}
where $G(t)$ denotes the normalized Gaussian instrument response. Among them, the slow component amplitude decreases gradually with temperature (Fig. 1(d)), while its lifetime shows a divergent tendency well below $T_{\mathrm{CDW}}$ (Fig. 1(e)). This abrupt modification of ultrafast relaxation dynamics prior to the critical temperature has been similarly observed in other CDW systems\cite{Lin2020,Li2022}. Concurrently, the oscillatory mode softens from 0.3 THz with rising temperature (Fig. 1(c)). Although undetectable above 120 K, extrapolation indicates its frequency approaches zero near 138 K. The low frequency and softening behavior proximate to $T_{\mathrm{CDW}}$ identify this oscillation as the AM. Notably, neither the AM frequency nor quasiparticle dynamics displays discernible anomalies near $T_N$, suggesting weak coupling between the CDW and magnetic orders.  
                                                                                                                                                                                                                                                                                                                                                                                                                                                                                                                                                                                                                                                                                                                                                                                                                                                                                                                                                                                                                                                                                                                                                                                                                                                                                                                                                                                                                                                                                                                                                                                                                                                                                                                                                                                                                                                                                                                                                                                                                                                                                                                                                                                                                                                                                                                                                                                                                                                                                    
\begin{figure*}[t]
 \centering
 \centering\includegraphics[width=15 cm]{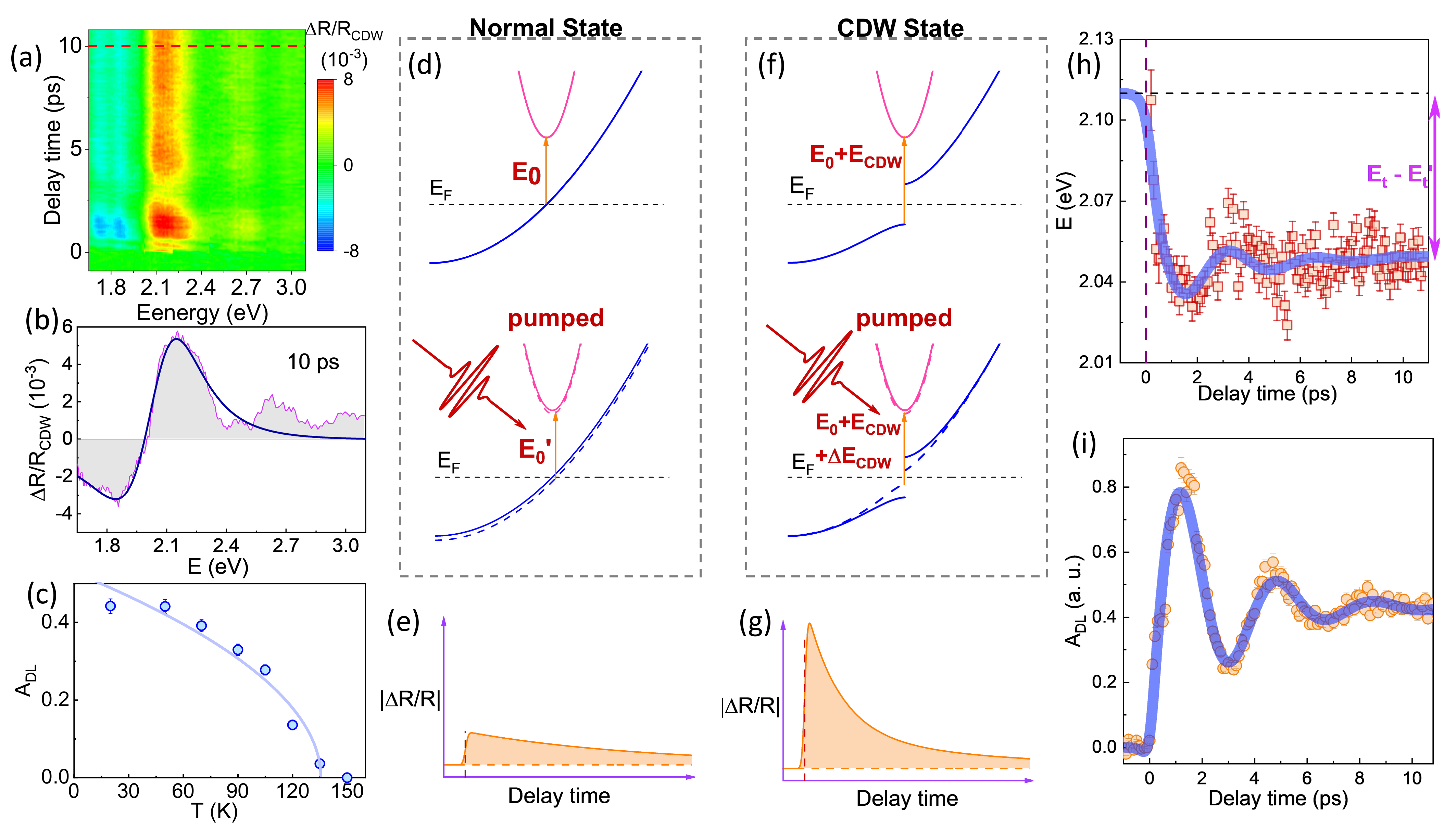}\\
 \caption{
\textbf{The CDW-specified interband transition suppressed by ultrafast laser.}  (a) The difference in broadband transient response between 20 K and 150 K, named as $\Delta R/R_{CDW}$. (b) $\Delta R/R_{CDW}$ spectra at 10 ps, where the blue curve is fitting result with Eq. (3).  (c) The amplitude parameter A$_{DL}$  extracted from $\Delta R/R_{CDW}$ at different low temperatures, and the blue curve represents a fit to an order-parameter-like temperature dependence ($\sqrt{1 - \frac{T}{T_c}}$). (d) In the normal state, there is no gap at the Fermi surface, allowing an upward transition with energy $E_0$, which is slight affected by the pump. (e) With the slight modulation on band structure, only a minor dynamics component with long lifetime can be observed. (f) Upon entering the CDW state, a gap of size $E_{CDW}$ opens, and the energy for the upward transition becomes $E_0$ + $E_{CDW}$. The magnitude of the bandgap opening is exaggerated here for clarity. With photoexcitation, the gap undergoes a variation of $\Delta E_{CDW}$ for its limited stability. (g) Due to the pump-induced transient gap restrain, the corresponding $\Delta R/R$ can record an extra pronounced fast component. (h) The pump-induced suppression of  transition energy $E_t$ by Eq. (3). (i) Time evolution of the A$_{DL}$. The dark blue curves are phenomenological fitting results as guide to the eyes.} \label{Fig:3}
\end{figure*}

As a typical collective excitation of the CDW, AM oscillations produce periodic fluctuations in ionic displacements. These fluctuations modify the CDW gap via electron-phonon coupling (EPC), altering the condensation energy and thereby generating an effective restoring force. In the long-wavelength limit, the AM frequency ($\omega_A$) satisfies the relation\cite{Chen2017,Gruner1994}:
\begin{equation}
\begin{aligned}
    \omega_A = \sqrt{\lambda\omega_{2k_F}^2} . \\
\end{aligned}
\end{equation}
where $\omega_{2k_F}$ denotes the softened phonon frequency in the normal state; and $\lambda$ represent the EPC strength. Using the $\omega_{2k_F}$ value of 1.3 meV at 400 K measured by IXS\cite{Korshunov2024}, we estimate an EPC strength of 0.9 for EuAl$_4$. This substantial EPC magnitude underscores its critical role in CDW formation\cite{Eiter2013,Zhu2015,Weber2011,Arguello2015}, consistent with prior studies\cite{Wang2024,Sukhanov2025}.

 Then we employ broadband ultrafast spectroscopy to capture the frequency- and time-domain evolution of transient reflectivity ($\Delta R/R (\omega, t)$) across 1.65–3.1 eV (Fig. 2(a)). At 20 K, the $\Delta R/R$ spectrum (Fig. 2(b)) exhibits two distinct components: (i) two negative dips at 1.9 eV and 2.3 eV near 0 ps with short lifetimes, attributed to excited carriers; and (ii) a subsequent slow evolution with complex energy distribution emerging after 1 ps delay (Fig. 2(c) shows representative spectra at 0 ps and 10 ps). The superimposed AM oscillations are faintly visible. In the normal state (Fig. 2(d)), while two negative dips near 0 ps survive, the delayed $\Delta R/R$ spectrum weakens significantly (Fig. 2(e)) with altered peak-valley characteristics. These characters are invisible in steady-state spectra(Section VII in \cite{Wu2026}).

Phenomenologically, the $\Delta R/R$ spectral response can be described by a derivative Lorentzian (DL) function as follows \cite{Aspnes1973,Jnawali2020}:
 \begin{equation}
\begin{aligned}
    \frac{\Delta R}{R}(E) =\sum_{j=1}^{n}Re[A_j e^{i\phi_j}(E-E_{t j}+i\Gamma_j)^{-2}]. \\
\end{aligned}
\end{equation}
Here, $n$ denotes the number of relevant interband transitions, $A_j$ and $\phi_j$ represent the amplitude and phase, $E_{tj}$ is the characteristic transition energy, and $\Gamma_j$ is the broadening coefficient. From this equation, the transition energies at specific delay times can be determined\cite{Jnawali2020}. While multiple components are needed to accurately reproduce the full $\Delta R/R$ spectra, doing so is mathematically infeasible owing to the large number of parameters involved.

Dual-color pump-probe measurements reveal that most relaxation components are temperature-independent (Section IV in \cite{Wu2026}), with only one component emerging in the CDW state. To isolate this CDW-specific contribution, we compute the difference between the transient response at 20 K and 150 K, denoted as $\Delta R/R$$_{CDW}$. As shown in Fig. 3(a), $\Delta R/R$$_{CDW}$ exhibits a negative response (light blue) below 2 eV and a positive feature (red) in the 2–2.3 eV range, while remaining near zero elsewhere (green). Focusing on the 10 ps delay—where coherent oscillations have fully damped under current pump fluence—the spectrum (Fig. 3(b)) is well reproduced by a single DL component with transition energy $E_t \approx$ 2 eV. The corresponding amplitude $A_\text{DL}$ (equivalent to $A_j$ in Eq. (3)) follows an order parameter-like temperature dependence below 135 K (Fig. 3(c)). This result unambiguously attributes the 2 eV optical transition to the CDW modulation. The underlying mechanism involves either excitations from occupied states below the Fermi level to the upper edge of the CDW gap, or from its lower edge to unoccupied states above the Fermi level. Band-structure analysis for EuAl$_4$ further reveal that such transitions are allowed (Section IX in \cite{Wu2026}).

The reflectivity is linked to the effective transition rate via the complex photoconductivity; in the perturbative limit, the amplitude of the transient reflectivity modulation is directly proportional to the variation in the transition rate. The effective transition rate $\Gamma$ between two specific electronic states can be expressed as\cite{Dressel2002}:
\begin{equation}
    \Gamma = W_0\, \rho \, f_i\left[ 1 - f_f \right],
\end{equation}
where $W_0$ represents microscopic transition matrix element, $\rho$ is the joint density of states, and $f_i$ and $f_f$ represent the occupation probabilities of the initial and final states, respectively.

Upon photoexcitation, the total variation of the effective transition rate yields a differential form comprising three distinct physical contributions:
\begin{equation}
    d\Gamma = \frac{\partial \Gamma}{\partial E} \, dE 
            + \frac{\partial \Gamma}{\partial f_f} \, df_f 
            + \frac{\partial \Gamma}{\partial f_i} \, df_i,
\end{equation}
where $E$ denotes the energy of the band portion associated with $W_0$. The second term (final-state Pauli blocking) and the third term (initial-state distribution modification) are ubiquitous across various photoexcited systems. In contrast, the magnitude of the first term (photo-induced band renormalization) depends critically on the robustness of the underlying electronic order. The normal state exhibits high resilience, remaining largely unperturbed under pump fluences up to $\sim$\,mJ/cm$^{2}$ (Figs. 3(d)); consequently, the first term contributes minimally, displaying only a weak, long-lived thermal signal (Fig. 3(e)). Conversely, the CDW order parameter is exceptionally fragile under photoexcitation. Even at identical fluences, it undergoes substantial modifications of its electronic structure (Fig. 3(f)). As a result, the photoinduced modulation of the CDW gap can reach several or even tens of meV. For instance, in EuAl$_4$, the transition energy E$_t$ drops from about 2.11 to 2.05 eV (equivalent to $dE$ $\approx$ 60 meV) within the first 1 ps, followed by a slow recovery accompanied by pronounced AM oscillations(Fig. 3(h)). This profound band renormalization gives rise to a prominent ultrafast component (Fig. 3(g)). Note that the schematic in Fig. 3(e,g) omits the $\sim$100-fs fast component, as it is unrelated to the phase transition dynamics.

\begin{figure}[t]
 \centering
 \centering\includegraphics[width=8.3 cm]{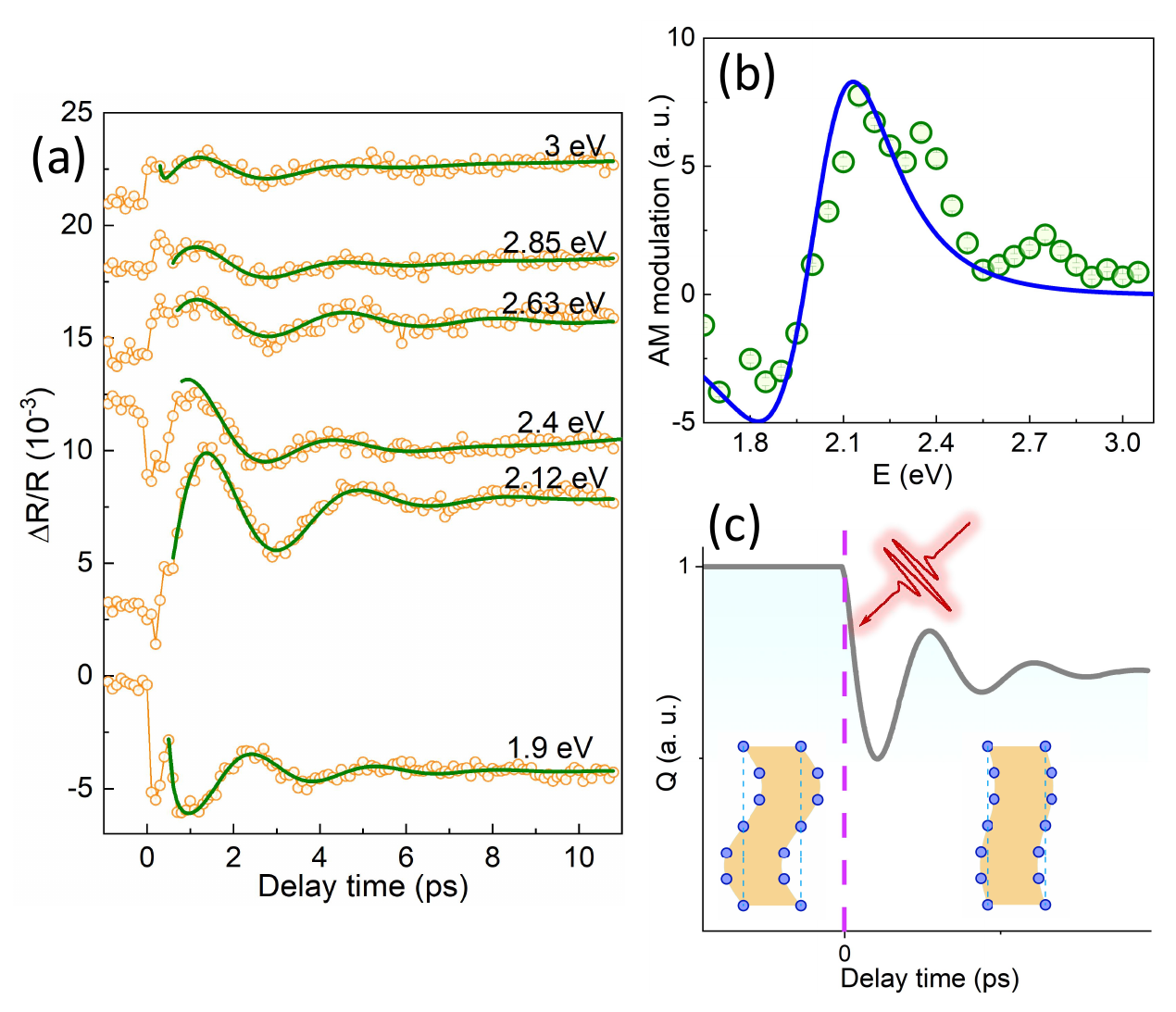}\\
 \caption{
\textbf{The AM modulated reflectance.} (a) The $\Delta R/R$ time traces at selected probe energy. The AM oscillations can be clearly seen, which shows significantly energy dependence. The green curves are waveform fitting to extract the AM amplitude. (b) Energy dependence of AM modulation on reflectance, which are obtained from fitting shown in panel (a). The blue curve is the DL line shape used in Fig. 3(b). (c) Schematic illustration of the time evolution of the normal coordinate of the CDW mode $ Q$. The bottom inset depicts the CDW distortion before and after photoexcitation, highlighting the transient suppression of the ordered state.} \label{Fig:4}
\end{figure}

Building upon the observation of coherent AM oscillations in the time-resolved $\Delta R/R$ signal [Fig. 3(a)], we now examine the energy dependence of the AM modulation amplitude similar to previous work\cite{Mann2016}. Figure 4(a) shows the extracted AM component at several probe energies, revealing a pronounced spectral shape that closely mirrors the differential CDW response $\Delta R/R_{\mathrm{CDW}}$. Quantitatively, the AM modulation amplitude follows the single-turn DL lineshape that is virtually identical to $\Delta R/R_{\mathrm{CDW}}$ [Fig. 4(b)]. This one-to-one correspondence demonstrates that both the photoinduced DL feature and the coherent AM oscillation originate from the same underlying mechanism: photoexcitation induced ionic displacements that modulate the CDW gap ($\Delta Q$), which are governed by the same energy-dependent coupling coefficient $\alpha(E)$(Section X in \cite{Wu2026}). Since the overall amplitude of $\Delta R/R_{CDW} (E)$ (equivalent to $A_{DL}$) scales with $\Delta Q$, by tracking the evolution of \( A_{\mathrm{DL}} \) at different pump–probe delays as shown in Fig. 3(i), we gain direct insight into the temporal dynamics of CDW-related ionic displacements and even the CDW formation in EuAl$_4$(Section XI in \cite{Wu2026})), which is a pivotal research topic in CDW systems\cite{Peierls1955,Woll1962,Varma1983,Valla2004}. Given that ultrafast excitation typically suppresses CDW distortion, $\Delta Q$ must be negative. We therefore reconstruct the trajectory of normal coordinate of the CDW mode (Fig. 4(c)): the CDW distortion weakens within  about 1 ps, accompanied by coherent AM oscillations, and settles into a long-lived state with reduced intensity.

The enhancement or emergence of an relaxation component upon gap opening is ubiquitous across superconductors\cite{Demsar1999,Chia2010}, CDW systems\cite{Wang2021,Yusupov2008}, heavy Fermion systems\cite{Zhang2017,Liu2020}, and even magnetically ordered materials\cite{Chu2017,Sun2021}, indicating that it originates from the generic formation of an energy gap rather than any specific microscopic mechanism. However, while the underlying physics is universal, its experimental manifestations vary significantly across different systems. Specifically, the gap opening may lead to a global sign reversal of transient reflectivity, a dramatic signal enhancement, or merely a weak relaxation tail, depending on the material details. This diversity arises from the interplay between intrinsic material properties and extrinsic probe conditions. Intrinsically, the spectral profile of the gap-related relaxation amplitude---exemplified by $\alpha(E)$ in EuAl$_4$ and analogous response functions in other ordered states---is shaped by the gap-induced reconstruction of the electronic density of states and the competing quasiparticle relaxation channels. Extrinsically, most ultrafast experiments employ a monochromatic 800-nm (1.55-eV) probe, whose spectral detuning relative to the characteristic transition energy $E_t$ at the gap edge varies substantially across systems. When the probe is far off-resonance with $E_t$, the gap-related component becomes negligible or unresolved; under near-resonant conditions, the coupling to gap-edge transitions is strongly enhanced. Furthermore, the sign of the transient signal is determined by the probe energy's position relative to the zero crossing of this material-specific spectral response function. Since the zero crossing shifts sensitively with the gap magnitude and carrier concentration, even modest doping can push a given system across this spectral node, triggering an apparent sign reversal that reflects the continuous evolution of the underlying electronic structure rather than a qualitative change in the ordered state. Although distinct band structures across different systems may shift analogous experimental signatures to varying energy scales, the underlying physical picture revealed in this work is expected to be universal.

In summary, we conduct a comprehensive investigation of non-equilibrium ultrafast dynamics in the CDW material EuAl$_4$. In addition to the AM with strong EPC, we observe a transient reflectance feature at around 2 eV in CDW state. We confirm that it originates from transitions associated with the CDW gap edge. Upon photoexcitation, the CDW gap shrinks, modulating the joint density of state and transition probability of high-energy transitions, ultimately leading to changes in the reflectivity spectrum. Our work provides deep insights into gap-modulated nonequilibrium dynamics, a quintessential ultrafast phenomenon, thereby filling a critical gap in the physical understanding of ultrafast spectroscopy.

\begin{acknowledgments}
This work was supported by the National Natural Science Foundation of China (No.  12404167, 12488201) and the National Key Research and Development Program of China (No. 2024YFA1408701, 2022YFA1403901).
\end{acknowledgments}

\bibliography{EuAl4}

\end{document}